\documentstyle[multicol,prl,aps,epsf]{revtex}
\begin{document}
\draft
\title{Exchange interactions and magnetic properties of the layered vanadates CaV$_2
$O$_5$, MgV$_2$O$_5$, CaV$_3$O$_7$ and CaV$_4$O$_9$}
\author{M. A. Korotin$^{(1)}$, I. S. Elfimov$^{(1)}$, V. I. Anisimov$^{(1)}$, M.
Troyer$^{(2)}$ and D. I. Khomskii$^{(3)}$}
\address{$^{(1)}$Institute of Metal Physics, Ekaterinburg GSP-170, Russia \\
$^{(2)}$Theoretische Physik, ETH Z\"urich, CH-8093 Z\"urich, 
Switzerland \\
$^{(3)}$Solid State Physics Dept of the Materials Science Center,\\
University of Groningen, 
Nijenborgh 4, 9747 AG Groningen, The
Netherlands}
\date{\today}
\maketitle

\begin{abstract}
We have performed {\it ab-initio} calculations of exchange couplings in the
layered vanadates ${\rm CaV}_2{\rm O}_5$, ${\rm MgV}_2{\rm O}_5$, ${\rm CaV}%
_3{\rm O}_7$ and ${\rm CaV}_4{\rm O}_9$. The uniform susceptibility of the
Heisenberg model with these exchange couplings is then calculated by quantum
Monte Carlo method; it agrees well with the experimental measurements. 
Based on our results we naturally explain the unusual magnetic properties of these
materials, especially the huge difference in spin gap between ${\rm CaV}_2%
{\rm O}_5$ and ${\rm MgV}_2{\rm O}_5$, the unusual long range order in ${\rm %
CaV}_3{\rm O}_7$ and the "plaquette resonating valence bond (RVB)" spin gap
in ${\rm CaV}_4{\rm O}_9$.
\end{abstract}

\begin{multicols}{2}

The series of insulating vanadates with the composition 
MV$_n$O$_{2n+1}$ is of considerable interest because it shows a 
variety of strange magnetic phenomena which are not completely 
understood.  Among these quasi-two dimensional layered materials 
${\rm CaV}_4{\rm O}_9$ has originally attracted a lot of attention 
as the first two-dimensional material with a gap in the spin 
excitation spectrum \cite{cav4o9exp} attributed to a plaquette RVB 
state \cite{cav4o9a}.  The other members of this family are also of 
interest.  ${\rm CaV}_3{\rm O}_7$ shows long range N\'eel order with 
an unusual ordering \cite{cav3o7exp} that could not be explained by 
simple models \cite{kontani}.  ${\rm CaV}_2{\rm O}_5$ has a large spin 
gap of about 600 K \cite{cav2o5}, but for ${\rm MgV}_2{\rm O}_5$ on 
the other hand, which has nearly the same lattice structure, there is 
only contradicting evidence for a possibly gapped state with a 
much smaller gap of only about 20 K \cite{mgv2o5exp}.

Despite extensive theoretical work on these 
materials\cite{cav4o9a,kontani,cav4o9b,cav4o9,cav4o9d,pickett,mila,miyahara} 
their magnetic properties are not yet fully understood.  In contrast 
to planar cuprates, 
where a hole in $x^2-y^2$-orbitals of Cu results in a strong 
antiferromagnetic exchange coupling for 180$^\circ$ bonds and a weak 
ferromagnetic one for 90$^\circ$ bonds, the interactions in these 
vanadates are much more complicated.  Not even the sign of many of the 
exchange couplings is obvious in these materials.  Only straightforward 
numerical {\it ab-initio} calculations can give information about 
relative and absolute values of the exchange couplings in these 
systems.

In this Letter we report on such an {\it ab-initio} calculation using 
the LDA+U method \cite{ci1} to compute the electronic structure and 
from it the exchange couplings.  We then use the quantum Monte Carlo 
(QMC) method to calculate the uniform susceptibility assuming these 
exchange couplings, and compare it with experimental measurements.  We 
find good agreement and are thus confident that our explanation of the 
magnetic properties of these materials  
captures the relevant physics.

The LDA+U method was shown to give good results for insulating 
transition metal oxides with a partially filled d-shell \cite{review}.  
The exchange interaction parameters can be calculated using a 
procedure based on the Greens function method which was developed by 
A.I.~Lichtenstein \cite{lichtexchange,lichtan}.  This method was 
successfully applied to calculate the exchange couplings in 
KCuF$_3$ \cite{lichtan} and in layered cuprates \cite{preprint}.

The LDA+U method \cite{ci1,review} is essentially the Local Density
Approximation (LDA) modified  by a potential correction restoring a
proper description of the Coulomb interaction between localized d-electrons
of transition metal ions. This is written in the form of a projection
operator:
\begin{equation}
\widehat{H}=\widehat{H}_{LSDA}+
\sum_{mm^{\prime }}\mid inlm\sigma \rangle
V_{mm^{\prime }}^\sigma \langle inlm^{\prime }\sigma \mid
\label{hamilt}
\end{equation}
\begin{eqnarray}
V_{mm^{\prime }}^\sigma &=&\sum_{\{m\}}
\{U_{m,m^{\prime \prime }
m^{\prime },m^{\prime \prime \prime }}
n_{m^{\prime \prime }
m^{\prime \prime \prime }}^{-\sigma } 
+(U_{m,m^{\prime \prime }
m^{\prime },m^{\prime \prime \prime }}\nonumber   
 \\
&&-U_{m,m^{\prime \prime }
m^{\prime \prime \prime },
m^{\prime }} )
n_{m^{\prime \prime }m^{\prime \prime
\prime }}^\sigma \} 
-U(N-\frac 12)+J(N^{\sigma}-\frac 12)\nonumber
\end{eqnarray}
where $\mid inlm\sigma \rangle $ ($i$ denotes the site, $n$ the main 
quantum number, $l$- orbital quantum number, $m$- magnetic number and 
$\sigma$- spin index) are d-orbitals of transition metal ions.  The 
density matrix is defined by:
\begin{equation}
n_{mm^{\prime }}^\sigma =-\frac 1\pi \int^{E_F}ImG_{inlm,inlm^
{\prime}}^\sigma(E)dE ,
\label{Occ}
\end{equation}
where $G_{inlm,inlm^{^{\prime }}}^\sigma (E)= \langle inlm\sigma \mid 
(E-\widehat{H})^{-1}\mid inlm^{^{\prime }}\sigma \rangle $ are the 
elements of the Green function matrix, $N^\sigma 
=Tr(n_{mm^{\prime}}^\sigma )$ and $N=N^{\uparrow }+N^{\downarrow }.$ 
$U$ and $J$ are screened Coulomb and exchange parameters. 
The 
$U_{mm^{\prime}m^{\prime\prime}m^{\prime\prime\prime}}$
is the screened Coulomb interaction among the $nl$ electrons 
which can be expressed via integrals over complex spherical harmonics 
and $U$ and $J$ parameters.  For the CaV$_n$O$_{2n+1}$ compounds the 
values of these parameters were calculated to be $U$=3.6~eV, 
$J$=0.88~eV   
via the so-called "supercell" procedure \cite{superlsda}.
The calculation scheme was realized in the framework of 
the Linear Muffin-Tin Orbitals (LMTO) method \cite{lmto} based on the 
Stuttgart TBLMTO-47 computer code.

The inter-site exchange couplings were calculated with a formula which 
was derived using the Green function method as second derivative of 
the ground state energy with respect to the magnetic moment rotation 
angle \cite{lichtexchange,lichtan}:
\begin{equation}
\label{exchange}
J_{ij}=\sum_{\{m\}}I_{mm^{\prime }}^i
\chi _{mm^{\prime }m^{\prime \prime}
m^{\prime \prime \prime }}^{ij}
I_{m^{\prime \prime }m^{\prime \prime \prime }}^j
\end{equation}
where the spin-dependent potentials $I$ are expressed in terms of the
potentials of Eq. (\ref{hamilt}),
\begin{equation}  \label{magpot}
I_{mm^{\prime }}^i=V_{mm^{\prime }}^{i\uparrow }-V_{mm^{\prime
}}^{i\downarrow }.
\end{equation}
The effective inter-sublattice susceptibilities are defined in terms of 
the LDA+U eigenfunctions $\psi $ as
\begin{equation}
\label{suscep}
\chi _{mm^{\prime }m^{\prime \prime }m^{\prime \prime \prime }}^{ij}=\sum_{
{\bf knn}^{\prime }}\frac{n_{n{\bf k\uparrow }}-n_{n^{\prime }{\bf
k\downarrow }}}{\epsilon _{n{\bf k\uparrow }}-\epsilon _{n^{\prime }{\bf
k\downarrow }}}\psi _{n{\bf k\uparrow }}^{ilm^{*}}\psi _{n{\bf k\uparrow }
}^{jlm^{\prime \prime }}\psi _{n^{\prime }{\bf k\downarrow }}^{ilm^{\prime
}}\psi _{n^{\prime }{\bf k\downarrow }}^{jlm^{\prime \prime \prime *}}.\nonumber
\end{equation}
The main building block of the crystal structures of the 
CaV$_n$O$_{2n+1}$ series is a  V ion  roughly in the center of a 
pyramid of oxygen ions.  
The 
lowest energy orbital is the V$3d$-orbital of $xy$-symmetry (using a 
convention where the axes of the coordinates system are directed 
toward the oxygen ions in the plane), which is the orbital whose lobes 
look at the directions, where the overlap with the oxygen is the 
smallest.  The crystal field splitting between $xy$-orbital and other 
$3d$-orbitals is so strong, that in the LDA band structure the $xy$-band 
is separated from the rest of V$3d$-bands by a small energy gap 
\cite{pickett}.
\begin{figure}
\epsfxsize=85mm
\centerline{ \epsffile{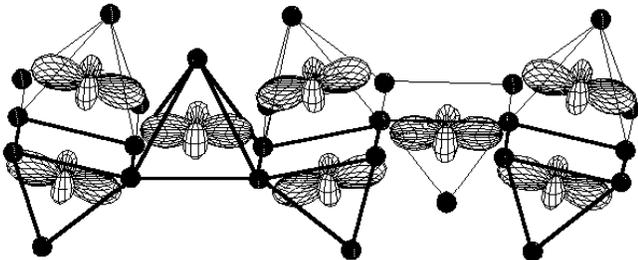} }
\narrowtext
\caption{The occupied d-orbitals of V$^{4+}$ ions in CaV$_3$O$_7$.
Oxygen atoms (denoted as small black circles) form pyramids
with V atoms inside them.}
\label{orbitals}
\end{figure}
As a consequence the degeneracy of the V$3d$-shell is lifted and the 
single $d$ electron of V$^{4+}$ ion occupies this $xy$-orbital, which 
reminds us of the cuprates, with a single hole in the 
$x^2-y^2$-orbital.  The important difference is that while in cuprates 
all copper atoms are in the same $(x,y)$-plane as the 
$x^2-y^2$-orbital, in these vanadates the vertices of the pyramids 
point alternatingly up and down with respect to the basal plane.  Thus 
the V ions in their centers are correspondingly above and below the 
central plane, as can be seen in Fig.~\ref{orbitals}.  As the 
$xy$-orbitals are parallel to this plane, the overlap (and hence the 
exchange couplings) are expected be stronger for vanadium ions 
situated on the same side of the plane.  We will show that this is 
indeed the case.  In addition to this alternation, a tilting of the 
pyramids is present in the crystal structure of these compounds, 
which seriously influences the interactions.

Another important difference with the cuprates is that the 
$xy$-orbital has a $\pi$-overlap with the in-plane oxygen atoms in contrast 
to a much stronger $\sigma$-overlap in case of $Cu^{2+}$.  Consequently 
one can expect much weaker exchange interaction in vanadates as 
compared to cuprates.  All the more surprising is thus the fact that 
the spin gap in CaV$_2$O$_5$ (616K \cite{cav2o5}), is {\it larger} 
than the typical values for the similar cuprate ladders ($\approx$ 460 K
 \cite{culadders}).

\begin{figure}[tbp]
\epsfxsize=180mm
\centerline{ \epsffile{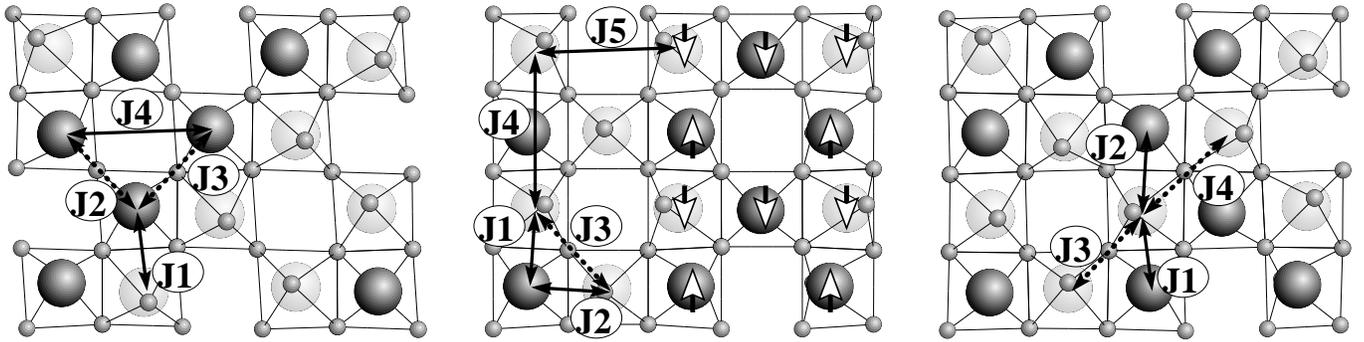} }
\caption{The basic crystal structure and the notation of exchange couplings
in CaV$_2$O$_5$ and MgV$_2$O$_5$ (left panel), CaV$_3$O$_7$ (middle panel)
and CaV$_4$O$_9$ (right panel). V atoms represented by large circles with
different colors have different $z$-coordinate. Oxygen atoms are shown by
small circles. The long range magnetic structure of CaV$_3$O$_7$ is depicted
by white arrows.}
\label{J_definition}
\end{figure}

Equation (\ref{exchange}) was derived as a second derivative of the 
total energy with respect to the angle between spin directions of the 
LDA+U solution.  The LDA+U method is the analogue of the Hartree-Fock 
(mean-field) approximation for a degenerate Hubbard model 
\cite{review}.  While in the multi-orbital case a mean-field 
approximation gives reasonably good estimates for the total energy, for 
the non-degenerate Hubbard model it is known to underestimate the 
triplet-singlet energy difference (and thus the value of the effective 
exchange coupling $J_{ij}$) by a factor of two for a two-site problem 
($E_{HF}=\frac{2t^2}{U}$ and $E_{exact} =\frac{4t^2}{U}$, where $t\ll 
U$ is inter-site hopping parameter).  In the problem under 
consideration there are two types of contributions to the exchange 
interaction parameters $J_{ij}$.  The first one is due to the $xy-xy$ 
orbitals hopping, and as only this orbital is half-filled this 
contribution directly corresponds to the non-degenerate Hubbard model 
and its value must be multiplied by a factor of two to correct the 
Hartree-Fock value.  Other contributions are due to the hoppings to 
all other orbitals and as the mean-field approximation is much better 
for multi-orbital model this part can be used unmodified.
\begin{table}
\label{valJ}
\caption{Calculated exchange coupling parameters (K).
"Minus" sign means ferromagnetic exchange.}
\begin{tabular}{lcccc}
\multicolumn{1}{|l}{ } &
\multicolumn{1}{c|}{CaV$_2$O$_5$} &
\multicolumn{1}{c|}{MgV$_2$O$_5$} &
\multicolumn{1}{c|}{CaV$_3$O$_7$} &
\multicolumn{1}{c|}{CaV$_4$O$_9$} \\
\hline
\multicolumn{1}{|l}{$J1$}&
\multicolumn{1}{c|}{--28} &
\multicolumn{1}{c|}{60} &
\multicolumn{1}{c|}{ 46} &
\multicolumn{1}{c|}{ 62}\\
\multicolumn{1}{|l}{$J2$}&
\multicolumn{1}{c|}{608} &
\multicolumn{1}{c|}{92} &
\multicolumn{1}{c|}{ --14} &
\multicolumn{1}{c|}{ 89}\\
\multicolumn{1}{|l}{$J3$}&
\multicolumn{1}{c|}{122} &
\multicolumn{1}{c|}{144} &
\multicolumn{1}{c|}{ 75} &
\multicolumn{1}{c|}{ 148}\\
\multicolumn{1}{|l}{$J4$}&
\multicolumn{1}{c|}{20} &
\multicolumn{1}{c|}{19} &
\multicolumn{1}{c|}{ 18} &
\multicolumn{1}{c|}{ 91}\\
\multicolumn{1}{|l}{$J5$}&
\multicolumn{1}{c|}{ } &
\multicolumn{1}{c|}{ } &
\multicolumn{1}{c|}{ 5} &
\multicolumn{1}{c|}{ }\\
\end{tabular}
\end{table}
In the crystal structure of these MV$_n$O$_{2n+1}$ compounds layers 
are formed by the VO$_5$ pyramids sharing edges of their bases.  The 
V atoms of the layer form roughly square lattice which is 
$1/(n+1)$-depleted (Fig.~\ref{J_definition}).  As mentioned earlier, 
the strongest interaction must be between V atoms which are situated 
on the same side of the plane (above or below) and are marked by the 
light gray and the dark gray circles in Fig.~\ref{J_definition}.
\end{multicols}
\widetext
\begin{multicols}{2}
In the MV$_2$O$_5$ crystal structure these atoms form ladders with 
interactions along the rung and the leg of the ladder denoted as $J_2$ 
and $J_3$ and interaction between ladders as $J_1$ (the notations are 
chosen to reflect the inter-atomic distances; the shortest one is 
between atoms on different sides of the plane).  In CaV$_3$O$_7$ the 
atoms on the same side of the plane form zigzag chains with 
interaction inside this chain denoted as $J_3$ and interactions 
between chains as $J_1$ and $J_2$.  For CaV$_4$O$_9$ such atoms form 
"metaplaquettes" (this term was proposed by W.Pickett \cite{pickett} 
to distinguish them from the plaquettes formed by atoms with the 
shortest V-V distance).  The interaction inside the metaplaquette is 
denoted by $J_3$, between metaplaquettes on the same side of the plane 
as $J_4$, and between metaplaquettes on the opposite sides of the 
plane by $J_1$ and $J_2$.

Our calculated values of the exchange couplings are presented in the 
Table I. It can immediately be seen that indeed the strongest 
interactions are between atoms on the same side of the plane ( the 
ladder exchanges $J_2$, $J_3$ for CaV$_2$O$_5$ and MgV$_2$O$_5$, the 
zigzag exchange $J_3$ for CaV$_3$O$_7$ and the metaplaquette exchange 
$J_3$ for CaV$_4$O$_9$).  However all other parameters are not 
negligible and for CaV$_4$O$_9$ the frustrating interactions between 
metaplaquettes are not less than 60\% of the value of interaction 
inside the metaplaquette.

For a comparison of these exchange couplings to measurements on the 
materials and to discuss the magnetic properties we consider the 
temperature dependence of the uniform susceptibility $\chi(T)$.  This 
quantity depends very sensitively on the exchange constants and can be 
both easily measured in experiments and calculated numerically for the 
Heisenberg model.  To calculate the $\chi(T)$ we use the continuous 
time version of the quantum Monte Carlo (QMC) loop algorithm 
\cite{loop}.  This algorithm uses no discretization of the imaginary 
time direction and the only source of systematic errors are thus 
finite size effects.  The lattice sizes were chosen large enough so 
that these errors are much smaller than the statistical errors of the 
QMC simulations.  All of these simulations, with the exception of the 
model for ${\rm CaV}_3{\rm O}_7$ suffer from the negative sign problem 
due to frustration effects.  Improved estimators \cite{ammon} can be 
used to lessen this sign problem, as was done in Ref.  \cite{miyahara} 
for similar models.  The frustration is especially strong in ${\rm 
MgV}_2{\rm O}_5$ and ${\rm CaV}_4{\rm O}_9$.  In these two cases 
$\chi(T)$ could not be calculated to temperatures far below the 
susceptibility maximum.  However $\chi(T)$ at intermediate 
temperatures is already very sensitive to the values of the exchange 
couplings and a comparison is thus possible.
\begin{figure}
\epsfxsize=85mm
\centerline{ \epsffile{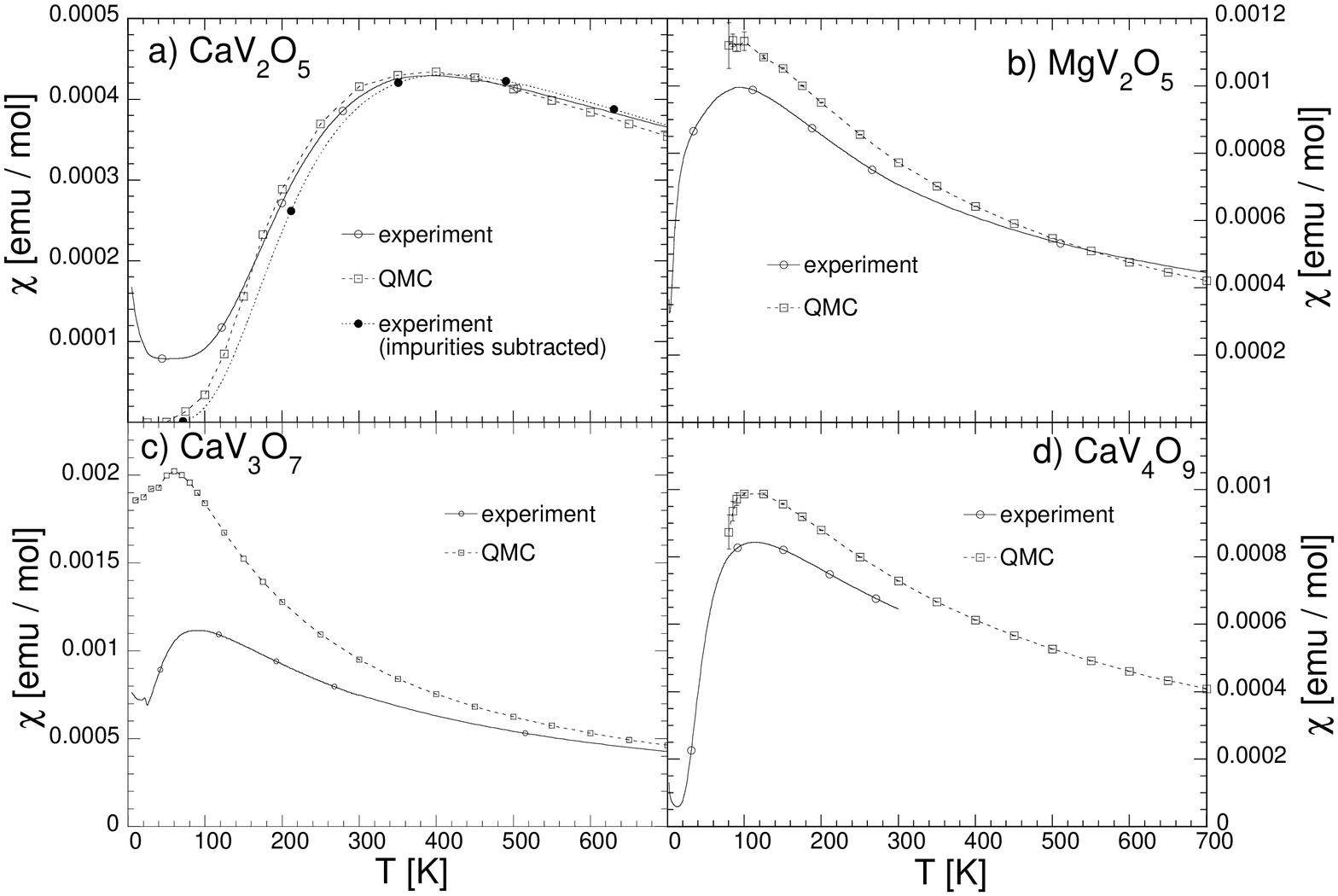} }
\narrowtext
\caption{Comparison of the temperature dependence of the uniform
susceptibility $\chi(T)$ measured in experiments
to that calculated in
QMC using the LDA+U values  
for  the exchange coupling of 
a) CaV$_2$O$_5$,
b) MgV$_2$O$_5$, 
c) CaV$_3$O$_7$ and 
d) CaV$_4$O$_9$.  
The $g$-factor of ${\rm V}^{+4}$ was 
taken to be $g=1.96$ .}
\label{fig:chi} 
\end{figure}
The $g$-factors for V$^{4+}$ ($d^1$,$S$=1/2) compounds are observed to 
be in a narrow range about $g = 1.96$ \cite{cvog}.  We use this value 
to compare (in Fig.  \ref{fig:chi}) the calculated $\chi(T)$ to 
measurements done by Isobe and Ueda \cite{chiexp} over a wide 
temperature range.  Taking note that the purity of the samples, and 
thus the normalization of the experimental data, is not known 
precisely and  keeping in mind how sensitive is $\chi(T)$ 
dependence to the values of the exchange couplings we find reasonable 
agreement of the calculated $\chi(T)$ with the experiments.

The agreement is especially good in the case of ${\rm CaV}_2{\rm O}_5$ 
where the exchange coupling $J_2$ on the rung of the ladder is 
dominant.  Our LDA+U results thus confirm the weakly coupled dimer 
picture proposed for this compound in \cite{cav2o5,miyahara}.

The magnitude of the exchange coupling however depends strongly on the 
tilting of the oxygen pyramid.  The compound MgV$_2$O$_5$ has nearly 
the same crystal structure as CaV$_2$O$_5$, but as the Mg ion has a 
smaller ionic radius than Ca ion, the tilting of the oxygen pyramids 
in MgV$_2$O$_5$ is stronger.  This causes the exchange coupling on the
 rung $J_2$ to be more than four times 
smaller and explains the huge difference in 
the uniform susceptibilities and spin gaps of these two compounds.

The exchange couplings in MgV$_2$O$_5$ are all of the same 
order, which positions this material in the strongly frustrated region 
of the trellis lattice phase diagram \cite{bruce}, precisely the 
region about which not much is known yet.  Even the experimental 
evidence is ambiguous, with different claims about the existence or 
non-existence of a small spin gap reported in the literature 
\cite{mgv2o5exp}.  More detailed experimental investigations are in 
progress, which will help to understand not only this compound but 
also the properties of the strongly frustrated trellis lattice
Heisenberg model.

Of the compounds under consideration only ${\rm CaV}_3{\rm O}_7$ has 
no frustration, and at low temperatures our QMC simulations gave a 
magnetically ordered N\'eel state, with the magnetic structure 
coinciding with the one observed in experiments.  This structure, with 
ferromagnetic order on the short 3-atom ``rungs'' and 
antiferromagnetic order along the infinite chains could not be easily 
explained previously.  The most surprising result is that the exchange 
couplings $J_{2}$ and $J_{1}$ which were thought to be 
equivalent\cite{kontani} are actually of opposite sign, most probably 
due to the tilting in the lattice structures (Fig.  1).  These 
exchange couplings also naturally lead to the observed experimental 
order, and no recourse to quantum fluctuation effects, as proposed by 
Kontani {\it et al.} \cite{kontani}, is necessary.  While there is 
qualitative agreement between the calculated and measured 
susceptibilities it is not as good as in the other compounds, due to 
closeness to a ferromagnetic state and large cancellation effects in 
the effective exchange couplings that increase the errors of the LDA+U 
exchange coupling estimates.

Of the four compounds ${\rm CaV}_4{\rm O}_9$ is the most studied one.  
We find that indeed the originally proposed plaquette-RVB state 
\cite{cav4o9a} is relevant, but with the larger meta-plaquettes 
($J_3$) being dominant, as suggested by Pickett \cite{pickett} and by 
Kodama {\it et al} \cite{cav4o9d} .  However, in contrast to the 
estimates of \cite{cav4o9d} we find that the coupling $J_{4}$, while 
smaller than $J_{3}$, is comparable to the other exchange couplings.

Our estimated exchange couplings can explain the order of magnitude of 
the spin gap of ${\rm CaV}_4{\rm O}_9$.  With our values of J's the 
unfrustrated $J_{3}$-$J_{4}$ model was shown to have a spin gap 
\cite{cav4o9a} of about 60K, which is further increased by frustration 
due to $J_{1}$ and $J_{2}$ \cite{cav4o9b}.  While exact theoretical 
estimates for the spin gap in this strongly frustrated regime are not 
available, this estimate agrees with the measured spin gap of 107K, 
and the good agreement between the measured and predicted uniform 
susceptibilities is a further indication for the validity of our 
estimates.

Additional comparisons can be done by calculating the magnon 
dispersion of the model and comparing it to experiments, as was 
recently done by Mambrini and Mila \cite{mila}.  According to their 
data, obtained on a small 16-site cluster our ratio $J_{3}/J_{4}$ 
might just be slightly too large to give the dispersion minimum at 
$(0,0)$, but further calculations on larger clusters are necessary to 
clarify this point.

To summarize, using LDA+U calculations of the exchange couplings for a 
series of layered vanadate compounds we can explain the puzzling 
magnetic properties of these materials and find good agreement between 
the {\it ab-initio} predictions for the uniform susceptibility and 
experimental measurements. 
 Such calculations are especially needed 
for systems such as the vanadates investigated here, where not only 
relative strength, but even the {\it sign} of the exchange 
interaction can not be estimated using simple geometrical arguments.
Applied to a series of layered vanadates we can explain the 
difference in spin gap of CaV$_2$O$_5$ and MgV$_2$O$_5$, obtain a 
natural explanation of the unusual magnetic order in CaV$_3$O$_7$
and confirm the plaquette-RVB state of CaV$_4$O$_9$ with {\it 
ab-initio} calculations.

We wish to thank W.E.~Pickett, A.I.~Lichtenstein, T.~Briere, 
D.C.~Johnston and K.~Ueda for stimulating discussions.  We are 
especially grateful to M.~Isobe and Y.~Ueda for providing us with 
their new unpublished susceptibility measurements.  The QMC 
calculations were performed using an automatically parallelizing C++ 
Monte Carlo library developed by one of the authors \cite{alea}.  This 
work was supported by the Russian Foundation for Basic Research 
(grants RFFI-98-02-17275 and RFFI-96-15-96598) , by the Netherlands 
Foundation for the Fundamental Study of Matter (FOM) and by the 
European Network OXSEN.

\end{multicols}

\end{document}